\documentclass[%
reprint,
 amsmath,amssymb,
 aps,
 prl,
]{revtex4-1}

\usepackage{graphicx}
\usepackage{dcolumn}
\usepackage{bm}

\begin{document}

\preprint{APS/123-QED}

\title{Evidence for the first-order phase transition\\ at the divergence region of activity expansions}

\author{M. V. Ushcats}
\affiliation{Taras Shevchenko National University of Kyiv, 2, Prosp. Academician Glushkov, Kyiv 03680, Ukraine}
\affiliation{Admiral Makarov National University of Shipbuilding, 
 9, Prosp. Heroes of Ukraine, Mykolayiv 54025, Ukraine 
}%
\author{L. A. Bulavin}
\affiliation{Taras Shevchenko National University of Kiev, 2, Prosp. Academician Glushkov, Kiev, 03680, Ukraine}


\date{\today}

\begin{abstract}
On the example of a lattice-gas model, a convincing confirmation is obtained for the direct relationship between the condensation phenomenon and divergent behavior of the virial expansions for pressure and density in powers of activity. The present study analytically proves the pressure equality for the low-density and high-density virial expansions in powers of density (in terms of irreducible cluster integrals or virial coefficients) exactly at the symmetrical points, where their isothermal bulk modulus vanishes, as well as for the corresponding expansions in powers of activity (in terms of reducible cluster integrals) at the same points (the points of their divergence). For lattice-gas models of arbitrary geometry and dimensions, a simple and general expression is derived for the phase-transition activity (the convergence radius of activity expansions) that, in particular, exactly matches the well-known phase-transition activity of the Lee -- Yang model. In addition, the study demonstrates that Mayer's expansion with the constant (volume independent) cluster integrals remains correct up to the condensation beginning, and the actual density-dependence may be taken into account for the high-order integrals only in more dense regimes beyond the saturation point.
\begin{description}
\item[PACS numbers] {05.20.-y; 05.20.Jj; 05.50.+q; 05.70.Ce; 05.70.Fh; 51.30.+i; 64.10.+h; 64.60.De; 64.70.F}

\end{description}
\end{abstract}

\maketitle


Since the 19th century, there have been many attempts to construct a strict and general theory for the gas-liquid phase transition on the basis of the Gibbs statistics \cite{Balescu, Pathria}, however the real success has been achieved only for two extremely specific statistical models: the mean-field approximation of intermolecular interactions (the van der Waals -- Maxwell equation of state \cite{Kac2, Lebowitz}) and two-dimensional lattice gas with the square-well potential (its condensation parameters were established by Lee and Yang \cite{LeeYang} on the basis of Onsager's solution of the Ising problem \cite{Onsager}). Unfortunately, more general approaches have failed in high-density regimes of fluids and, especially, at the phase-transition region \cite{Balescu}: the solution of the Ornstein -- Zernike equation \cite{OZ} vanishes in the transition regimes \cite{Martynov}, and various expansions for the partition function (or its logarithm) in powers of a small parameter \cite{Balescu} also have singularities in dense states. For example, the divergent behavior of the well-known virial expansions in powers of activity or density \cite{Balescu, Pathria,Mayer} remains an actual problem of statistical mechanics, mathematical physics, physical chemistry, etc. The situation was well described by the authors of the one-dimensional van der Waals -- Maxwell equation: "Many attempts, thus far unsuccessful, have been made to construct a rigorous theory of condensation phenomena from such expansions. In fact, we believe that such a construction is very difficult, if not \emph{impossible} ..." \cite{Kac2}.

Recent studies of Mayer's cluster expansion \cite{Mayer} for the partition function of realistic interaction models (i.e., the models that include intermolecular attraction as well as repulsion: the Lennard-Jones model \cite{LJ1,LJ2} and its modifications \cite{UPJ1,JCP2}, square-well \cite{KofkeSW}, Morse \cite{Morse}, Yukawa \cite{Yukawa} potentials, etc.) have renewed interest to the problem. In particular, an exact generating function for Mayer's expansion in terms of irreducible cluster integrals (virial coefficients) \cite{PRL, Bannur} has allowed deriving the equation of state (UEOS) beyond the adequacy region of the conventional virial expansions \cite{PRL}. This equation yields the constancy of pressure at any density beyond the point $\rho _G$, where the isothermal bulk modulus of the virial expansion in powers of density (virial equation of state or VEOS) vanishes \cite{PRE1,JCP1}, that, in turn, may indicate the beginning of the condensation process at the vicinity of this point. Consequent studies of Mayer's expansion in terms of reducible cluster integrals \cite{UPJ4,PRE4,Pramana} (equation of state in the parametrical form of expansions for pressure and density in powers of activity, AVEOS) have demonstrated its divergence exactly at the same point $\rho _G$, but the actual character of this divergence \cite{UPJ4,PRE4,Pramana,Gibbs2} agrees with the behavior of Mayer's expansion in terms of irreducible integrals (where the activity dependence is excluded) and corresponds to the known thermodynamic signs of the first-order phase transition. 

On the other hand, all the mentioned above approaches to Mayer's cluster expansion involve serious mathematical and technical limitations that make their results questionable. Despite the rapid development of modern computational techniques \cite{Kofke1,UPJ2,Kofke3}, in practice, the studied equations always include finite (or roughly approximated \cite{JCP3, UPJ3, Kofke5}) sets of cluster integrals (reducible as well as irreducible) that makes an adequate comparison with the experimental data almost impossible. Moreover, the numerical studies of any infinite series cannot be considered as mathematically correct, especially, in its divergence region. For a long time, the density, $\rho _G$, where the VEOS isothermal bulk modulus vanishes, used to be considered as a spinodal point (the boundary between metastable and absolutely instable states), because the isotherms of the VEOS, which includes only 3-5 power terms, have an interval similar to the famous van der Waals loop (the Maxwell construction is still widely used for all theoretical as well as empirical equations despite the fact that it has rigorously been proved only for a specific kind of the mean-field approximation \cite{Kac2,Lebowitz}). Although the observed behavior of the VEOS with the terms of higher orders completely differs from that of the van der Waals equation the destruction of such long-standing stereotypes remains a difficult problem that needs strict and persuasive arguments.

Here, Mayer's cluster-based method is applied to the lattice-gas model of arbitrary geometry and dimensions with regard for its recently established "hole-particle" symmetry. On the one hand, this approach provides a simple but convincing evidence for the condensation phenomenon at the AVEOS divergence region (thus confirming the results of the previous studies in terms of reducible or irreducible integrals) and, on the other hand, the obtained results absolutely agree with the exact Lee~-- Yang solution for a specific two-dimensional lattice-gas model.

For any classical system of interacting particles (including the lattice-gas model and, with some restrictions, even certain quantum systems \cite{Pathria, Kilpatrick}), Mayer's cluster expansion \cite{Mayer} represents the logarithm of the grand partition function (the pressure, $P$) and its derivative with respect to the chemical potential, $\mu$, (the particle number density, $\rho = N / V$) as the following series, 
\begin{equation}
\left. \begin{array}{l}
\frac{P}{{{k_B}T}} = \sum\limits_{n = 1}^\infty  {{b_n}{z^n}} \\
\rho  = \sum\limits_{n = 1}^\infty  {n{b_n}{z^n}} 
\end{array} \right\} \label{eq1} ,
\end{equation}
in terms of reducible cluster integrals $\left\{ b_n \right\} $ and powers of activity $z = {\lambda ^{ - 3}}\exp \left( {\mu } / {{k_B}T} \right) $ ($\lambda  = h/\sqrt {2\pi m{k_B}T} $ is the de Broglie wavelength).
In order to exclude the activity dependence, Eq.~(\ref{eq1}) (AVEOS) is often transformed to the well-known virial equation of state (VEOS) \cite{Mayer},
\begin{equation}
\frac{P}{{{k_B}T}} = \rho \left( {1 - \sum\limits_{k \ge 1} {\frac{k}{{k + 1}}{\beta _k}{\rho ^k}} } \right)\label{eq2},
\end{equation}
in terms of irreducible cluster integrals $\left\{ \beta _k \right\} $, which are related (in a complex manner \cite{Mayer}) to the reducible integrals, $\left\{ b_n \right\} $.
 
It was first shown by the Mayers \cite{Mayer} and later confirmed in other studies \cite{PRL, JCP1, PRE1, Bannur, PRE4} that the transformation of the AVEOS to VEOS (and, therefore, the validity of the VEOS itself) is strongly restricted at subcritical temperatures even in its convergence region: the density, $\rho$ in Eq.~(\ref{eq2}), must not exceed the value $\rho _G$ defined as the minimum real positive root of the following equation,
\begin{equation}
\sum\limits_{k \ge 1} {k{\beta _k}{\rho _G ^k}} = 1 \label{eq3},
\end{equation}
which corresponds to the vanishing of the VEOS isothermal bulk modulus [it is easy to check that Eq.~(\ref{eq3}) exactly defines all the points, where the derivative of pressure in Eq.~(\ref{eq2}) with respect to density vanishes].
 
In addition, the $\rho _G$ is directly related to the convergence radius, $z_G$, for the activity series in Eq.~(\ref{eq1}) (AVEOS) \cite{Mayer, JML, Bannur, PRE4}:
\begin{equation}
{z_G} = {\rho _G}\exp \left( { - \sum\limits_{k \ge 1} {{\beta _k}\rho _G^k} } \right) \label{eq4}.
\end{equation}

This activity, $z_G$, is exactly the divergence point of the activity series, $\sum {{n^2}{b_n}{z^n}} $ (and, therefore, the AVEOS series, $\sum {{n}{b_n}{z^n}} $, $\sum {{b_n}{z^n}} $, in accordance with the Cauchy-Hadamard theorem) \cite{Mayer, JML, Bannur, PRE4, Pramana}.

The vanishing of the isothermal bulk modulus at the point $\rho _G \left( z_G \right) $ means that the AVEOS divergence yields the jump of density at constant pressure and activity (chemical potential), and such behavior of the AVEOS has been confirmed for a number of statistical models of fluids \cite{UPJ4,PRE4,Pramana}. Moreover, the equation of state based on the exact generating function in terms of irreducible integrals (UEOS) \cite{PRL, Bannur} (there is no activity dependence and, hence, no divergence in the UEOS) really yields the pressure constancy at any density beyond the $\rho _G$.

For a wide range of lattice-gas models (the sole restriction is that the interaction potential must have a hard core identical to a lattice cell), the "hole-particle" symmetry of the Hamiltonian has recently been established \cite{PRE2}, and high-density equations of state have been derived \cite{PRE2, PRE3}, which are symmetrical to the AVEOS and VEOS: the expansions for pressure and density (SAVEOS),
\begin{equation}
\left. \begin{array}{l}
\frac{P}{{{k_B}T}} = {\rho _0}\left( {\frac{{{u_0}}}{{{k_B}T}} + \ln \frac{{{\rho _0}}}{\eta }} \right) + \sum\limits_{n \ge 1} {{b_n}{\eta ^n}} \\
\rho  = {\rho _0} - \sum\limits_{n \ge 1} {n{b_n}{\eta ^n}} 
\end{array} \right\}
,
\label{eq5} 
\end{equation}
and virial expansion for pressure (SVEOS),
\begin{eqnarray}
&\frac{P}{{{k_B}T}} = {\rho _0}\left[ {\frac{{{u_0}}}{{{k_B}T}} + \ln \left( {\frac{{{\rho _0}}}{{\rho '}}} \right) + \sum\limits_{k \ge 1} {{\beta _k}{{\rho '}^k}} } \right] \nonumber \\ 
&+ \rho '\left[ {1 - \sum\limits_{k \ge 1} {\frac{k}{{k + 1}}{\beta _k}{{\rho '}^k}} } \right], \label{eq6}
\end{eqnarray}
where 
\begin{equation}
\eta  = \frac{{\rho _0^2}}{z}\exp \left( {\frac{{2{u_0}}}{{{k_B}T}}} \right) \label{eq7}
\end{equation}
is the reciprocal activity; $\rho ' = \rho _0 - \rho $ is the "hole" number density; $\rho _0$ is the close-packing density; $u_0$ is the potential energy per particle in the close-packing state.

It is obvious that the applicability of Eqs.~(\ref{eq5}), (\ref{eq6}) (SAVEOS and SVEOS) is limited symmetrically to Eqs.~(\ref{eq1}), (\ref{eq2}) (AVEOS and VEOS). In the SVEOS, the "hole" number density, $\rho' $, must not exceed the value $\rho _G$ from Eq.~(\ref{eq3}), i.e., Eq.~(\ref{eq6}) remains adequate only in very dense regimes, when $\rho$ is not lower than 
\begin{equation}
{\rho _L} = {\rho _0} - {\rho _G}\ \label{eq8}
\end{equation}
that is the point where the isothermal bulk modulus of Eq.~(\ref{eq6}) (SVEOS) vanishes.

As to the SAVEOS [see Eq.~(\ref{eq5})], its radius of convergence, $\eta _L$, is also defined by Eq.~(\ref{eq4}) (i.e., $\eta _L \equiv z_G$), and its divergent behavior is symmetrical to that of the AVEOS: there is a jump of density from the $\rho _L$ to lower values of density at constant pressure and chemical potential \cite{UPJ4}.

In order to establish the relationship between the condensation phenomenon and symmetrical divergence of the AVEOS, SAVEOS, the numerical studies have recently been performed for the two-dimensional Lee -- Yang lattice-gas model \cite{UPJ4}. Unfortunately, the obtained results did not demonstrate the persuasive convergence to the exact Lee -- Yang solution: the equations with limited number of the known virial coefficients even yields the substantial difference of pressure at the symmetrical points of the phase-transition, $\rho _G \left( z_G \right)$ and $\rho _L \left( \eta _L \right)$.

Nevertheless, some simple but very important aspects of the "hole -- particle" symmetry have escaped attention in the previous studies. The equality of the convergence radiuses ($z_G \equiv \eta _L $) for the series of activity, $z$, in Eq.~(\ref{eq1}) and reciprocal activity, $\eta$, in Eq.~(\ref{eq5}) is obvious. However, there can only be one activity that is equal to its reciprocal quantity from Eq.~(\ref{eq7}):
\begin{equation}
z_G \equiv \eta _L = \rho _0 \exp \left( {\frac{{u_0}}{{k_B}T}} \right), \label{eq9}
\end{equation}

It is important that Eq.~(\ref{eq9}) makes the pressure, $P \left( z_G \right)$, in Eq.~(\ref{eq1}) \emph{identical} to that, $P \left( \eta _L \right)$, in Eq.~(\ref{eq5}), because 
\[
\left. \rho _0 \left({\frac{{u_0}}{{k_B}T}} + \ln {\frac{{\rho _0}}{\eta}} \right) \right|_{\eta = z_G} = 0.
\]

At supercritical temperatures [when Eq.~(\ref{eq3}) has no real root, Eq.~(\ref{eq4}) is invalid, both AVEOS and SAVEOS are always convergent], the activity, $z_G$ from Eq.~(\ref{eq9}), just defines the density, $\rho _0 / 2$, where Eqs.~(\ref{eq1}) and (\ref{eq5}) coincides.
 
\begin{figure}
\includegraphics{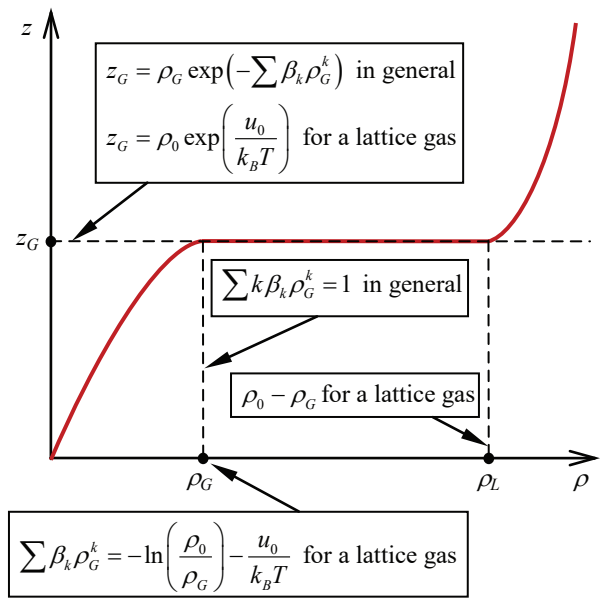}
\caption{\label{fig1}Typical (schematic) subcritical isotherm of a lattice gas (in the $\rho$ -- $z$ plane) that must theoretically be produced by the AVEOS (on the left of the $\rho _G$ point) and SAVEOS (on the right of the $\rho _L$ point) with complete and accurate sets of cluster integrals.}
\end{figure}

At subcritical temperatures, there are two distinct densities, $\rho _G$ and $\rho _L $ [the divergence points of AVEOS and SAVEOS, respectively: the first is defined by Eq.~(\ref{eq3}), and the second -- symmetrically in Eq.~(\ref{eq8})] for one value of activity [defined by Eq.~(\ref{eq9})] at one pressure in Eqs.~(\ref{eq1}), (\ref{eq5}). Therefore, the $z_G$ from Eq.~(\ref{eq9}) is \emph{the phase-transition activity}: the first-order phase transition must yield a \emph{density jump at constant pressure and chemical potential} (i.e., the constant activity, see Fig.~\ref{fig1}).

Indeed, this value, $z_G$ from Eq.~(9), exactly matches the phase-transition fugacity,
\[
y_G = \frac {z_G}{\rho _0} = \exp \left( { -\frac{4\epsilon}{{k_B}T} } \right),
\] 
of the Lee -- Yang solution \cite{LeeYang} (in their two-dimensional square model, each particle interacts only with its four nearest neighbors, and the interaction energy per particle, $u_0$, is $-4\epsilon$, when the depth of the potential well is $2\epsilon$). 

On the other hand, the definition of $z_G$ in Eq.~(\ref{eq9}) is \emph{more genera}l and must be valid for lattice-gas models of arbitrary geometry and dimensions with various (even anisotropic) interaction potentials: the geometry of the model as well as the characteristics of the potential are enclosed into the $u_0$ parameter.

An additional important fact concerns the equations in terms of irreducible integrals: VEOS and SVEOS. In accordance with Eq.~(\ref{eq9}), Eq.~(\ref{eq4}) is transformed to the following condition for the irreducible (virial) series of lattice-gas models at the points where the isothermal bulk modulus of both equations (\ref{eq2}) and (\ref{eq6}) vanishes,
\[
\sum\limits_{k \ge 1} {{\beta _k}{\rho _G ^k}} = {\ln \frac {\rho _G}{\rho _0}} - \frac {u_0}{{k_B}T} .
\]

The last condition not only \emph{equalizes the pressure of VEOS and SVEOS} at those points,
\[ 
P\left( \rho _G \right) \equiv P\left( \rho _L \right), 
\] 
(that makes it impossible to interpret the $\rho _G$ and $\rho _L $ in other way than binodal points) but can also be useful in further studies of the virial series.

In fact, all these features provide a much needed strict confirmation for some recently obtained results \cite{PRL,JCP1,JCP3,UPJ3,UPJ4,PRE4} concerning not only the lattice-gas model but other statistical models too, and this confirmation may finally end the long-standing disputes about the adequacy of Mayer's expansion, nature of its divergence, character of the $\rho _G$ point (i.e., whether it belongs to the binodal or spinodal), etc. The VEOS and AVEOS are general equations for various pair-wise interaction models, and, therefore, the fact that the density, $\rho _G$ from Eq.~(\ref{eq3}) (the zero point of the VEOS isothermal bulk modulus), is the saturation point for the lattice gas proves that it has the same sense for the other models with realistic interaction potentials. 

The difference may only be quantitative (due to the difference of irreducible sets for different interaction models) and not qualitative (at least for realistic interaction models that include both attraction and repulsion). The equality of pressure and chemical potential (activity) at the saturation and boiling points ($\rho _G$ and $\rho _L $, respectively) demonstrates that Eq.~(\ref{eq3}) really defines the binodal: $\rho _G$ cannot belong to the spinodal (as it was often considered earlier). The divergence region of the AVEOS is exactly bounded by the $\rho _G$, and this divergence indeed yields the jump of density at constant values of pressure and chemical potential (some modern approximations of infinite virial series \cite{JCP3, UPJ3, Kofke5} even yield a discontinuity of the isotherm tangent at the $\rho _G$ vicinity that makes the theoretical isotherms very similar to real ones).

Another important result is that Mayer's expansion with the constant (volume independent) cluster integrals remains correct up to the density $\rho _G$ (activity $z_G$). As a rule, the cluster integrals are defined in infinite limits, but, for the integrals of very high (macroscopic or thermodynamic) orders, this is not always correct: at very dense states, they must essentially depend on the real system volume, and their contribution also becomes determinative (this contribution even causes the AVEOS to diverge). In practice, the neglecting of such volume dependence for high-order cluster integrals produces the inadequate behavior of the AVEOS (as well as UEOS in terms of irreducible integrals) at very dense states \cite{PRL,PRE1,JCP1}: equations with constant cluster integrals actually yield the density jump to infinity (the essential discontinuity of density) instead of the finite jump to the boiling point $\rho _L $. In case of the discrete lattice-gas model, the usage of symmetrical equations (SAVEOS and SVEOS) very simply resolves this problem (see Fig.~\ref{fig1}), but, for continuous models of matter, the "hole-particle" symmetry is not so obvious. This issue still awaits a solution, however the present study directly confirms that \emph{the volume dependence of the cluster integrals may really be neglected at least in the regimes when} $\rho \le \rho _G$.

In conclusion, Mayer's cluster-based approach provides a strict statistical basis for the theoretical description of the condensation phenomenon. There is a general theoretical condition for the saturation point, $\rho _G$, [see Eq.~(\ref{eq3})]. The subcritical temperatures can be defined as those at which Eq.~(\ref{eq3}) has at least one positive root, $\rho _G$, and the divergent behavior of the density series in AVEOS (i.e., derivative of the partition function logarithm with respect to the chemical potential) provides a clear explanation for the condensation beginning at the $\rho _G$ density: increasing contribution of the high-order cluster integrals in the partition function statistically means the possibility to form liquid droplets (large clusters) in the gaseous phase.

It should additionally be noted that the presented results do not contradict those obtained in previous analytical \cite{LeeYang, Langer} and numerical \cite{Fonseca} studies of the lattice-gas model as well as Ising problem. For example, Lee and Yang \cite{LeeYang} considered a circle in the complex plane of activity where the grand partition function of a lattice gas vanishes, and they obtained the jump of density (in the present approach -- the divergence of the activity expansion for density) at the corresponding real $z$ which exactly matches $z_G$ in Eq.~(\ref{eq9}). In the Langer droplet model of a lattice gas \cite{Langer}, the same activity is treated as a divergence point of the cluster expansion and condensation point of the system.
 
Of course, the existing extremely limited data on the virial sets still do not allow an accurate quantitative determining of the condensation parameters [$\rho _G$ and $P \left(\rho _G \right)$] even for the simplest statistical models, but this problem may now be considered as technical rather than theoretical.

\nocite{*}

\providecommand{\noopsort}[1]{}\providecommand{\singleletter}[1]{#1}%
\end{document}